%% file: confpaper.tex
\documentclass[11pt]{article}

\usepackage{graphicx}
\usepackage{dcolumn}
\usepackage{amsmath}
\usepackage{units}
\usepackage{lineno}

\input{babarsym}

\newcommand{\BtoDs}{\mbox{$\Bbar^0\rightarrow D^{*+} \ell^- \bar{\nu}_\ell$}}
\newcommand{\mnusq}{\ifmmode{{{M}^2_\nu}} \else {${{ M}_{\nu}}^2$}\fi}

\newcommand{\BABARPubYear}    {07}

\newcommand{\BABARConfNumber} {008}
\newcommand{\SLACPubNumber} {12740}

\long\def\inst#1{\par\nobreak\kern 4pt\nobreak
    {\it #1}\par\vskip 10pt plus 3pt minus 3pt}

\begin{document}


{\pagestyle{empty}

\begin{flushleft}
\end{flushleft}

\begin{flushright}
\babar-CONF-\BABARPubYear/\BABARConfNumber \\
SLAC-PUB-\SLACPubNumber \\
August 2007 \\ 
\end{flushright}

\par\vskip 2cm

\begin{center}
\Large \bf \boldmath 
Measurement of the $\Bz\to X_{u}^- \ell^+ {\nu_\ell}~$ decays near the
kinematic endpoint of the lepton spectrum and 
search for violation of isospin symmetry 
\end{center}
\bigskip

\begin{center}
\large The \babar\ Collaboration\\
\mbox{ }\\
\today
\end{center}
\bigskip \bigskip

\begin{center}
\large \bf Abstract
\end{center}

We present the first measurement of the $\Bz\to X_{u}^- \ell^+ {\nu_\ell}~$ 
partial branching fraction in the end-point region of the lepton momentum spectrum, 
above the threshold for $B\ra X_c\ell {\nu_\ell}$ decays.
The analysis is based on a sample of 383 million $\Upsilon(4S)$ decays 
into \BB\ pairs collected with the \babar\ detector at the \pep2 $e^+e^-$
storage rings. We select $\BzBzb$ events by partially reconstructing one $\B$ meson
via the \BtoDs\ decays then select $\Bz\to X_{u}^- \ell^+ {\nu_\ell}~$ decays 
identifying a second high momentum lepton.
In the momentum interval ranging from $2.3$ to $2.6$~\gevc we measure the partial
branching fraction $\Delta{\cal B}(B^0\to X_u\ell\nu)=(1.30\pm0.21_{stat}\pm0.07_{syst})\times 10^{-4}$
where the first error is statistical and the second is systematic. 
By comparing this measurement with the one obtained from untagged $B$ decays we obtain 
$R^{+/0}=\Delta{\cal B}(B^0\to X_u\ell\nu)/\Delta{\cal B}(B^+\to X_u\ell\nu)= 1.18 \pm 0.35_{stat} \pm 0.17_{syst}$.
Using this measurement we extract a limit on the contributions from 
processes breaking isospin symmetry in charmless semileptonic $B$ decays.

\vfill
\begin{center}
Contributed to the 
XXIII$^{\rm rd}$ International Symposium on Lepton and Photon Interactions at High~Energies, 8/13 -- 8/18/2007, Daegu, Korea
\end{center}

\vspace{1.0cm}
\begin{center}
{\em Stanford Linear Accelerator Center, Stanford University, 
Stanford, CA 94309} \\ \vspace{0.1cm}\hrule\vspace{0.1cm}
Work supported in part by Department of Energy contract DE-AC03-76SF00515.
\end{center}

\newpage

\input{authors_lp2007.tex}

\newpage

\section{INTRODUCTION}

The precise determination of $|V_{ub}|$, the magnitude of the Cabibbo-Kobayashi-Maskawa \cite{CKM} 
matrix element, with well-understood uncertainties is one of the prime goals of 
heavy flavor physics. Recently, significant progress has been made with larger 
data samples available at the $B$-Factories and a variety of improved 
experimental techniques \cite{PDG_bob}. Advances in QCD calculations 
of leading and subleading contributions to partial decay rates 
in restricted regions of phase space, and the measurements of the $b$-quark mass and 
non-perturbative parameters from inclusive spectra in 
$B \to X_s \gamma$ and $B \to X_c \ell \nu$ 
\footnote{We indicate with $X$ the hadronic system
in semileptonic $B$ decays. We use the notation $X_u$ and $X_c$
when referring, respectively, to charmless and charmed hadronic system.} 
have resulted in much reduced errors on $|V_{ub}|$.  
One of the effects that is not included in current calculations 
of the partial decay rate, is weak annihilation (WA) \cite{bigi_wa}, which is expected to 
contribute at the level of a few percent \cite{neubert_wa,voloshin_wa,ligeti_wa,gambino_wa}. 
Simply speaking, WA refers to the annihilation 
of the $b-\ubar$ pair to a virtual $W$ boson, and results in an enhancement 
of the decay rate near the endpoint of the $q^2$ spectrum. Here $q^2$ refers to 
the mass squared of the virtual $W$.

Experimentally, WA should be observable as a violation of isospin invariance, 
i.e. difference in the partial decay rates of $B^0 \to \X_{u}^- \ell^+ \nu$ and 
$B^+ \to X_{u}^0 \ell^+ \nu$, at high $q^2$, since it 
occurs only for charged $B$ mesons. 

In this paper, we report a first measurement of the partial branching 
fraction for inclusive $B^0\to X_{u}^- \ell^+ \nu$ decays 
\footnote{ By charged lepton, $\ell$ we mean here only
  electron or muon. Charge conjugated states are always implied
  throughout this paper. Momentum and energy are computed in the 
\FourS\ frame, unless the lab frame is explicitly mentioned.}
above 2.3~\gevc of the charged lepton momentum. \BzBzb\ events 
produced at the \FourS\ resonance are tagged by the partially reconstructed  
$\BtoDs$ decays. We identify the charmless semileptonic decay of 
the second $B$ meson in the event and compare its partial decay rate with 
the partial rate for the sum of charged and neutral $B$ mesons previously published \cite{bad1047}, 
and extract the difference in these partial decay rates between $B^+$ and $B^0$ mesons.

\section{THE \babar\ DETECTOR AND DATASET}
We use a data sample of 383 million $B \bar B$ pairs
produced by the PEP-II asymmetric-energy $e^+e^-$ collider
and collected by the \babar\ experiment \cite{NIM} at the Stanford
Linear Accelerator Center on the \FourS\ resonance (on-resonance data), 
and about 36~\invfb collected 40~\mev below the resonance (off-resonance data) 
to study non-\BB\ (continuum) background events.
A detailed description of the \babar\ detector, of charged 
and neutral particle reconstruction and identification is provided
elsewhere \cite{NIM}.
Trajectories of charged particles are measured with two tracking
systems inside a 1.5-Tesla superconducting solenoid, a 5-layer
double-sided silicon vertex tracker and a 40-layer drift
chamber (DCH).
Both tracking systems are equipped to measure energy loss due to 
specific ionization ($dE/dx$), which is used to discriminate pions, kaons, electrons,
muons, and protons.
Additional particle identification is provided by Cerenkov 
radiation, which is generated in an array of silica
bars surrounding the DCH and is detected by an array of
phototubes.
The energy from electromagnetic showers in a CsI(Tl) crystal calorimeter
is measured and used to reconstruct photons and to identify
electrons.
The iron flux return of the solenoid is instrumented with layers
of resistive-plate chambers and limited-streamer tubes,
which are used to identify muons.
For background and efficiency corrections that cannot be measured from data, 
we use a full simulation of the detector based on GEANT4 \cite{geant}. 
The equivalent luminosity of the simulated \FourS$\to$ \BB event sample amounts 
to about 960~\invfb. 


\section{EVENT SELECTION}

The analysis proceeds in two steps. First \BB\ events are tagged searching for  
\BtoDs\ events identified with a partial reconstruction technique \cite{partial} 
described below. Second, in the tagged sample we 
identify $B^0\to\X_u\ell^+\nu$ decays searching for events with an additional 
high momentum lepton. 

We select \BtoDs\ events on the tag side, by partial reconstruction of the decay 
$D^{*+} \to \pi_{s} D^0$ using only the charged lepton from the 
\Bzb\ decay and the soft pion ($\pi_{s}$) from the $D^{*+}$ decay. 
The $D^0$ decay is not reconstructed, resulting in a high selection
efficiency. 
To suppress leptons from several background sources, the tag 
lepton must have a momentum in the range $ 1.4 < P_{\ell,tag} < 2.3 $
GeV/$c$. The momenta of the $\pi_{s}$ candidates must lie in the range 
$60 < p_{\pi_{s}} <200 \mevc$. To reduce continuum background,
we select events with at least 5 charged tracks with a  ratio of the second 
to the zeroth order Fox-Wolfram \cite{wolfram} moment $R_2 < 0.5$.
We construct a likelihood discriminator ${\cal L}$, using 
as input the lepton momentum $P_{\ell,tag}$, the $\pi_{s}$ momentum $p_{\pi}$ 
and the probability that $\ell_{tag}$
and the $\pi_s$ originate from a common vertex, constrained to the
beam-spot in the plane transverse to the beam direction.
We apply a cut on this discriminator to suppress background and, in case
of multiple $\ell_{tag}-\pi_{s}$ candidates, to select the candidate
with the highest likelihood value of ${\cal L}$.

We approximate the mass of the undetected neutrino 
\begin{equation}
\mnusq_{,tag} = \left( \frac{\sqrt{s}}{2} - {E}_{\dsp} - E_{\ell^-} \right)^2 - 
({\bf{P}}_{\dsp} + {\bf{P}}_{\ell^-} )^2, 
\label{eq1}
\end{equation}
where $\sqrt{s}/2$ is the beam energy in the \epem\ center-of-mass frame, 
$E_{\ell^-}$ and ${\bf{P}}_{\ell^-}$ are the energy and momentum vector of the lepton.
The energy ${E}_{\dsp}$ and the momentum ${\bf{P}}_{\dsp}$ of
the $\dsp$ meson are estimated as a linear function of the energy of the slow pion 
$\pi_{s}$, with parameters obtained from the simulation. We approximate the 
direction of the $\dsp$ to be that of the $\pi_{s}$.

The distribution of $\mnusq_{,tag}$ peaks at zero for 
$\Bz \rightarrow D^{*} \ell \bar\nu_{\ell} X$ events, while it extends over a wider range 
for all other events (see Fig.~\ref{f:mnu}). We define a signal region, 
$-3<\mnusq_{,tag}<2$ GeV$^2$/$c^4$, which contains $~98\%$ of the signal events, 
and a side band region, $-10<\mnusq_{,tag}<-4$ GeV$^2$/$c^4$, which is populated mostly
by background events.

Events that contribute to the peak at zero in the $\mnusq_{,tag}$ 
distribution, are mainly due to the following processes:
(a) $\Bzb \rightarrow D^{*+} \ell^{-} \bar\nu_{\ell}$ decays (primary),
(b) $\Bzb \rightarrow D^{*+}\Db, \Db\to\ell^{-}X$, $\Bzb \rightarrow D^{*+} \tau^{-} \bar\nu_{\ell}$, 
$\tau^-\rightarrow \ell^{-}X$ (cascade),
(c) $\Bzb \rightarrow D^{*+} h^-$ (fake), where the hadron 
($h = \pi,K)$ is erroneously identified as a lepton, (in most of the cases a muon), and 
(d) $\Bb \rightarrow D^{*+} (n \pi) \ell^- \bar{\nu}_{\ell}$, (with the number of $\pi$: $n\ge 1$), where the $D^{*+} (n \pi)$ may 
or may not originate from an excited charm state (\dstrstr). 
Source (a) accounts for about 90\% of the events that contribute to the peak in the 
$\mnusq_{,tag}$ distribution, while source (d) contains a sizeable contribution from
\Bp\  which constitutes the peaking background on the tag side, and sources (b) and (c)
account for few percent of the peak.  Non-peaking background is due to 
remaining \BB\-decays and due to continuum processes.

The continuum background is taken from the off-resonance data sample scaled by the luminosity 
ratio of the on-resonance and off-resonance data sets. 
The \BB\-background is taken from simulation.
We validate the simulation of the non-peaking background by comparing
on-resonance data with the sum of 
\BB\ Monte Carlo simulation and off-resonance data in a wrong-charge 
sample (WC), which is selected by requiring that the lepton 
and the soft pion have equal electrical charge \cite{prlDkpi}.
\begin{figure}[!htb]
\begin{tabular}{c c}
\hskip -1cm
\includegraphics[width=0.58\linewidth]{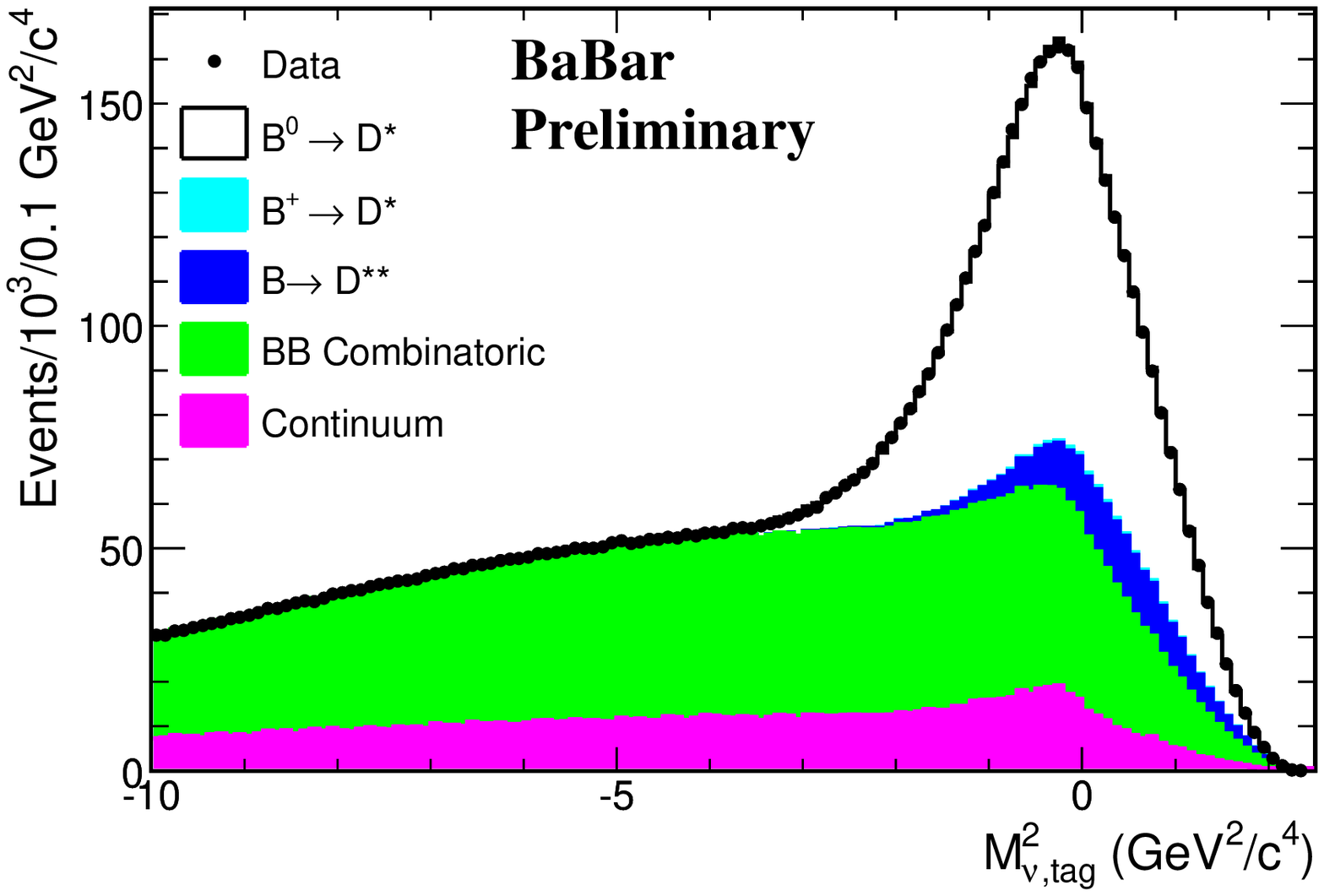}&
\hskip -1.3cm
\includegraphics[width=0.58\linewidth]{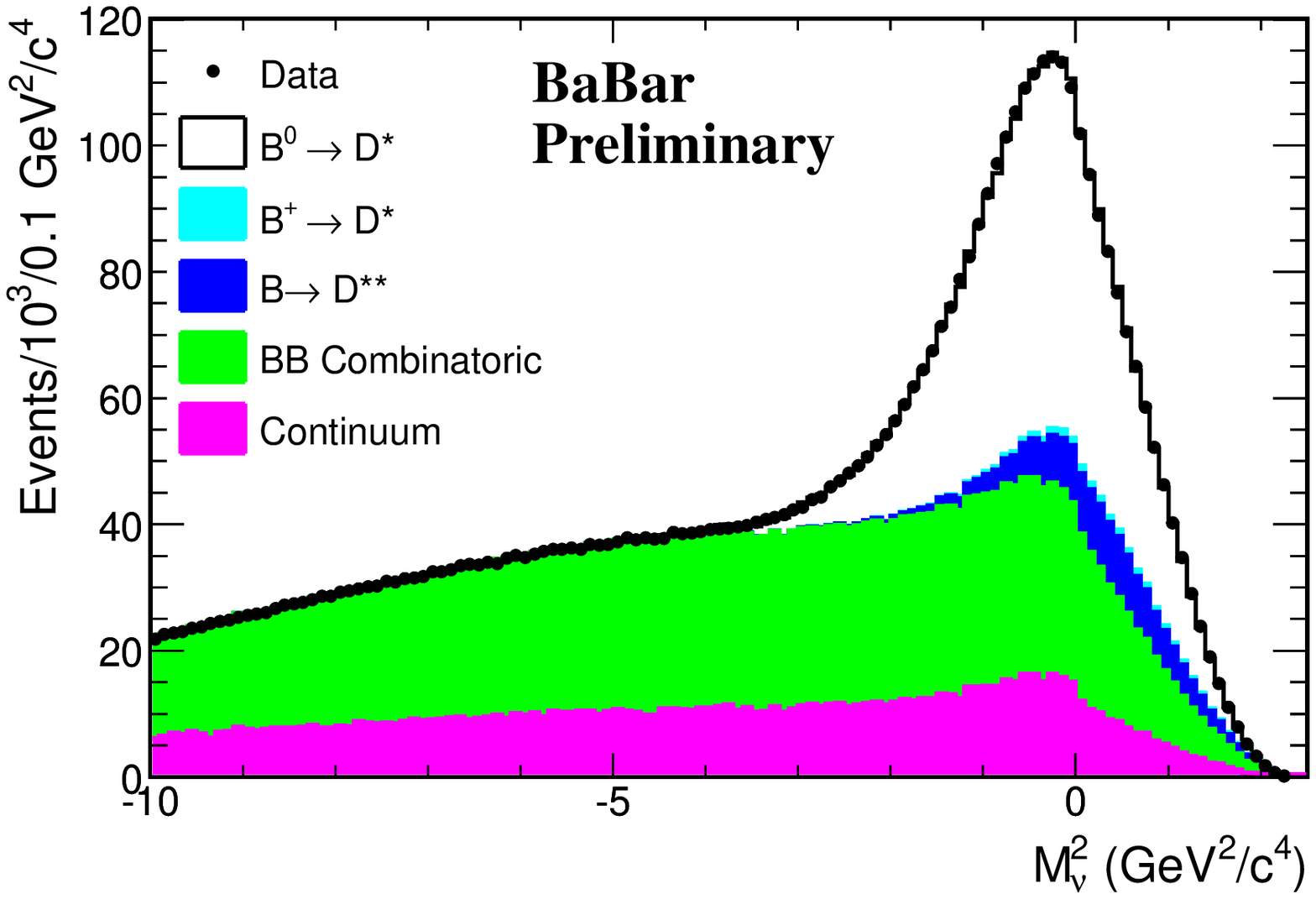}\\
\end{tabular}
\vspace*{-1.0cm}
\caption{The $\mnusq_{,tag}$ distribution of the tag sample summed over all the 
lepton momenta and all data-taking
periods, compared to the results of the fit shown as the sum of various 
contributions (stacked histograms); $\ell_{tag}=e$ (left) and $\ell_{tag}=\mu$ (right).} 
\label{f:mnu}
\end{figure}
We determine ${\cal N}_{\Bz}$, the number of tagged $\Bzb$ in our sample by a $\chi^2$
fit to the $\mnusq_{,tag}$ distribution in the interval
$-10.0<\mnusq_{,tag}<2.5$ GeV$^2$/$c^4$.
To reduce the sensitivity of the result to the details of the simulation, such as
the description of the $\pi_s$ reconstruction efficiency and the modeling of the
$B\to D^{*}\ell\nu$ and $B\to D^{**}\ell\nu$ decays,
we perform the fit in ten bins of the momentum, $P_{\ell,tag}$, of the tagged lepton.
The free parameters in the fit are the 
number of events from primary decays, from $B\to D^{**}$ decays, and from combinatorial
\BB background. 
The $\mnusq_{,tag}$ shapes of these samples are taken from simulation. 
The continuum contribution is taken from off-resonance data and 
fixed and the fractions of events from sources (b) and (c) 
are fixed to the prediction from the simulation. We divide the data into ten 
different subsamples, separating by lepton kind and different data-taking periods. 
We perform the fit for each of these subsamples. 
Figure~\ref{f:mnu} shows the result of the fit. 

According to isospin symmetry, the relative contributions of $\Bm$ and $\Bzb$ 
to the \dstrstr\ component should be 2/3 and 1/3, respectively.
The isospin symmetry hypothesis is validated 
with a precision of about 
 10$\%$ \cite{mixing}, so we assign (66$\pm$7)$\%$ of the $D^{**}$ events to the peaking $B^-$ background
and the rest to the \Bzb events. The fit shows that decays with a 
$D^{**}$ constitute 8$\%$ of the peaking 
sample. We assign a systematic uncertainty of $0.8\%$ to the fitted number of
\Bzb\ events due to isospin symmetry.

The total number of tagged $\Bz$ decays with $\mnusq_{,tag}>-10.0$ GeV$^2$/$c^4$ 
is ${\cal N}_{\Bz}=(3606.4\pm9.2_{stat}\pm47.1_{syst})\times 10^3$,
where the systematic error includes uncertainties on the $\mnusq_{,tag}$
shapes predicted by the simulation for each sample, on the fraction of
events from samples (b) and (c), and on the composition of the \dstrstr.


To enrich charmless semileptonic decays, we select tagged events with an
additional identified lepton (electron or muon) with momentum $P_{\ell}>2.2\gevc$. 
In this momentum range the identification efficiency is
about $0.95$ for electrons and $0.60$ for muons. 

Continuum production is the largest source of background at high lepton
momentum ($P_{\ell}>2.4 \gevc$). We therefore apply some further
selection criteria, which depend on the flavor of the two charged leptons 
($ee,~ \mu\mu,$ or $e\mu$).
These criteria are optimized using off-resonance events and on-resonance events
well above the kinematic limit for $ B\rightarrow X_u\ell \nu_\ell$
decays, $P_{\ell}>2.8~\gevc$. 

We reject events if the angle between the charged leptons
is close to zero or $\pi$, or if the invariant mass of these two leptons is less than $0.5~\gev/c^2$.  
To reduce the number of radiative Bhabha
events, we ask for at least six charged tracks in $ee$ events. 
We require the aplanarity $A$ of the event, defined in Ref.\cite{jetset}, 
to exceed 0.002. 
 
The missing momentum $p_{miss}$ is computed in the \FourS\ frame, as the vector sum
of the momenta of all charged tracks and calorimeter showers.
We select events with ${\bf p}_{miss}$ pointing inside the detector 
acceptance 
and  in the range $0.5 < |{\bf p}_{miss}| < 3.8 \gevc$. 
These criteria reduce the continuum background by a factor of 7.7 (3.5), 
and retain $74\%$ ($83\%$) of the $e$ ($\mu$) signal events. 

We reduce $B\to J/\psi,\psi'\to\ell\ell$ background by rejecting 
$e^\pm e_{tag}^\mp$ or $\mu^\pm \mu_{tag}^\mp$
pairs if their invariant mass is consistent
with the $J/\psi$ or $\psi$' mass. We also 
pair $\mu^{\pm}$ with tracks of opposite charge and 
reject them if the invariant mass of the pair is consistent with a $J/\psi$ or
$\psi$'.

We reduce background from $B \rightarrow X_c \ell \nu_\ell$ decays by
rejecting events with two or more kaons (either $K^{\pm}$ and $K_s$). 
We also reject events if the charged lepton combined with a low momentum
$\pi$ of opposite charge forms a second $D^{*-}$.

In the interval $2.3 < P_{\ell} < 2.6$ \gevc, 
the selection criteria retain  $25~(20) \%$ of the 
$B \rightarrow X_c\ell \nu_\ell$  background, 
and $70\% ~(65\%)$ of the $e (\mu)$ signal. 


\section{SIGNAL YIELDS} 

We group correctly tagged $\Bzb$ decays on the basis of the associated lepton:
\begin{enumerate}
\item {\bf Signal}, $\ell$ comes from $\Bz\to X_u\ell^+\nu_\ell$ decay.
\item{\bf \Bzb background}, where $\ell$ comes from either: 
\begin{enumerate}
  \item $B^0\to\X_c \ell \nu_\ell$ decays ($c$-background) ;
  \item $B^0\to D h X$ with the hadron $h$ misidentified as a lepton
    (fake) ;
  \item secondary leptons from $B^0\to D\to\ell X$, $B^0\to\tau\to\ell X$ 
and $B^0\to \psi\to\ell X$ decays (cascade) 
\end{enumerate}

\end{enumerate} 

We determine the number of signal events as a subsample of the tagged events 
with an extended binned maximum likelihood fit to the
$\mnusq_{,tag}$ distribution, using a fit method that properly accounts
for the statistical uncertainties of both the data and the simulation \cite{barlow}.
We perform three fits for three partially overlapping intervals of
$\Delta P_\ell$: $2.2-2.6~$\gevc, $2.3-2.6~$\gevc, and $2.4-2.6~$\gevc.

We fit the $\mnusq_{,tag}$ data distribution with the sum of three distributions:
combinatoric background (sum of continuum and non-peaking \BB\ ), 
peaking \BB\ background, and $B^0\to X_u\ell^+\nu$ signal. 
The peaking \BB\ background is mostly due to $B \rightarrow
X_c \ell \nu$ decays. Its amount is fixed to the
Monte Carlo prediction, adjusted to the latest measurements of 
semileptonic branching fractions and form-factor parameters. 
The $\mnusq_{,tag}$ shape for combinatorial background is taken from 
the wrong-charge data control sample.
 
The results of the three fits are detailed in Table~\ref{t:final}. Figure~\ref{f:mnufit}
shows the comparison between the fit results and the data in the intervals 
$\Delta P_\ell$ = 2.2-2.6 \gevc (top)
and 2.3-2.6 \gevc (middle) and 2.4-2.6 \gevc (bottom), separately for $e$ and $\mu$.

\begin{table*}
\begin{center}
\caption{\label{t:final} Event yields for $e$ and $\mu$ with 
$\mnusq_{,tag}>-3~$GeV$^2/$c$^4$, for three intervals of lepton momentum. 
Only the statistical errors are reported. 
$\Delta {\cal B}^0$ is the partial branching fraction in units of $10^{-4}$, 
separately for $e$ and $\mu$. 
The last row shows the final result obtained from their averages accounting 
for the systematic errors.} 
\begin{small}
\begin{tabular}{r|c|c|c|c|c|c} 
\hline
$\Delta P_\ell$&\multicolumn{2}{c|}{$2.2-2.6\gevc$} &\multicolumn{2}{c|}{$2.3-2.6\gevc$}&\multicolumn{2}{c}{$2.4-2.6\gevc$}\\  
\hline 
                     & $e$		& $\mu$ & 	$e$		&	$\mu$ 			& $e$	& $\mu$\\
\hline \hline
Data 	 	     &1051	&1073 & ~~452	& ~~428	& ~219	&~177 \\
\hline\hline
$P(\chi^2)$          & 0.06                  & 0.87 &0.10                   & 0.93                  & 0.13                  & 0.78 \\
\hline
$N^{comb}$	     &463$\pm$21 	&352$\pm$17& 242$\pm$17	& 156$\pm$12 	& 112$\pm$11 	&60$\pm$7\\
\hline
$B^-_{tag}$           &61.1$\pm$4.4		&69.3$\pm$4.5	& 18.0$\pm$2.4		& 25.2$\pm$2.7		&7.8$\pm$1.6		&8.5$\pm$1.5\\
Cascade		     &32.7$\pm$3.5		&23.5$\pm$2.7	& 17.6$\pm$2.7		& 14.3$\pm$2.1		&9.1$\pm$2.1		&7.0$\pm$1.5\\
Fake $\ell$	     &3.2$\pm$1.4		&95.0$\pm$5.6 	& 1.7$\pm$1.0		& 53.7$\pm$4.2		&1.3$\pm$0.9		&16.7$\pm$2.3\\
$D^{**}\ell\nu$	     &0.5$\pm$0.4		&0.7$\pm$0.5 	& 0			&0			&0			&0 \\
$D^{*}\ell\nu$	     &151.5$\pm$6.9 		&152.1$\pm$6.8 	& 12.7$\pm$1.9		&12.9$\pm$2.0		&1.0$\pm$0.6 		&0.2$\pm$0.2 \\
$D\ell\nu$	     &77.5$\pm$4.9		&83.0$\pm$5.1 	& 14.8$\pm$2.1		&11.6$\pm$1.8		&0.6$\pm$0.5		&0.2$\pm$0.2 \\
\hline
$X_u\ell\nu$	     &250.9$\pm$51.8         &339.4$\pm$50.2 & 131.1$\pm$36.9        & 172.0$\pm$32.2        &77.4$\pm$26.6          &96.9$\pm$21.8  \\
\hline\hline
$\varepsilon(\Delta P_\ell)~\%$ & 35.9$\pm$1.2 & 28.5$\pm$1.2 & 37.6$\pm$1.7 & 32.0$\pm$1.5 & 39.0$\pm$2.5 & 30.3$\pm$2.1 \\
1+$\delta_{rad}$  & 1.0903 & 1.0228 & 1.1026 & 1.0302 & 1.1195 & 1.0379 \\
$\Delta {\cal B}^0$ & 2.11$\pm$0.44 & 3.40$\pm$0.50  & 1.07$\pm$0.30 &  1.53$\pm$0.29 &  0.61$\pm$0.21 & 0.92$\pm$0.21 \\ 
\hline
Average $\Delta {\cal B}^0$ & \multicolumn{2}{c|}{2.62$\pm$0.33$\pm$0.16} & \multicolumn{2}{c|}{1.30$\pm$0.21$\pm$0.07}& \multicolumn{2}{c}{0.76$\pm$0.15$\pm$0.05}\\  
\hline
\end{tabular}
\end{small}
\end{center}
\end{table*}

\begin{figure*}
\begin{tabular}{c c}
\includegraphics[width=8cm]{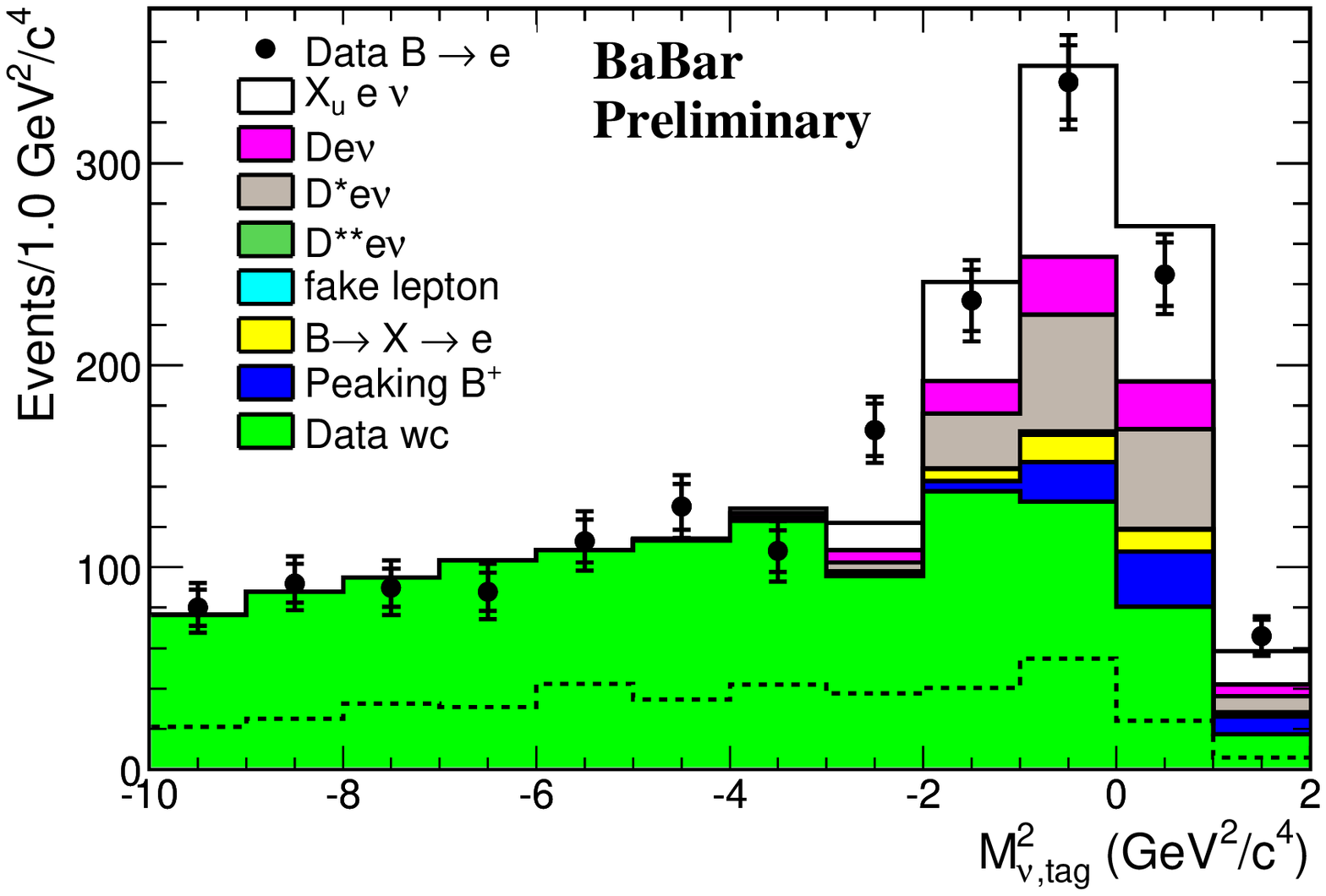}&
\includegraphics[width=8cm]{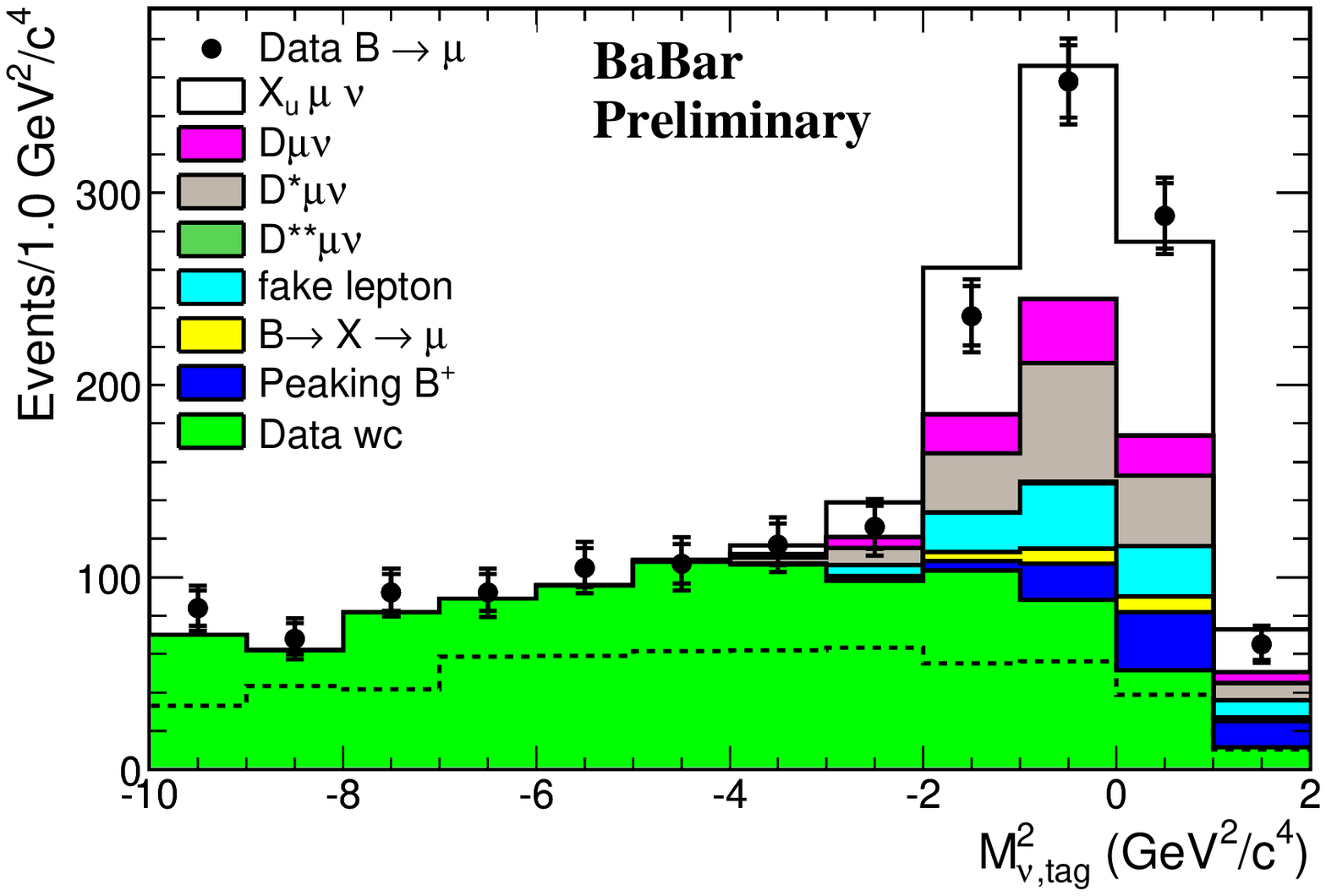}\\
\includegraphics[width=8cm]{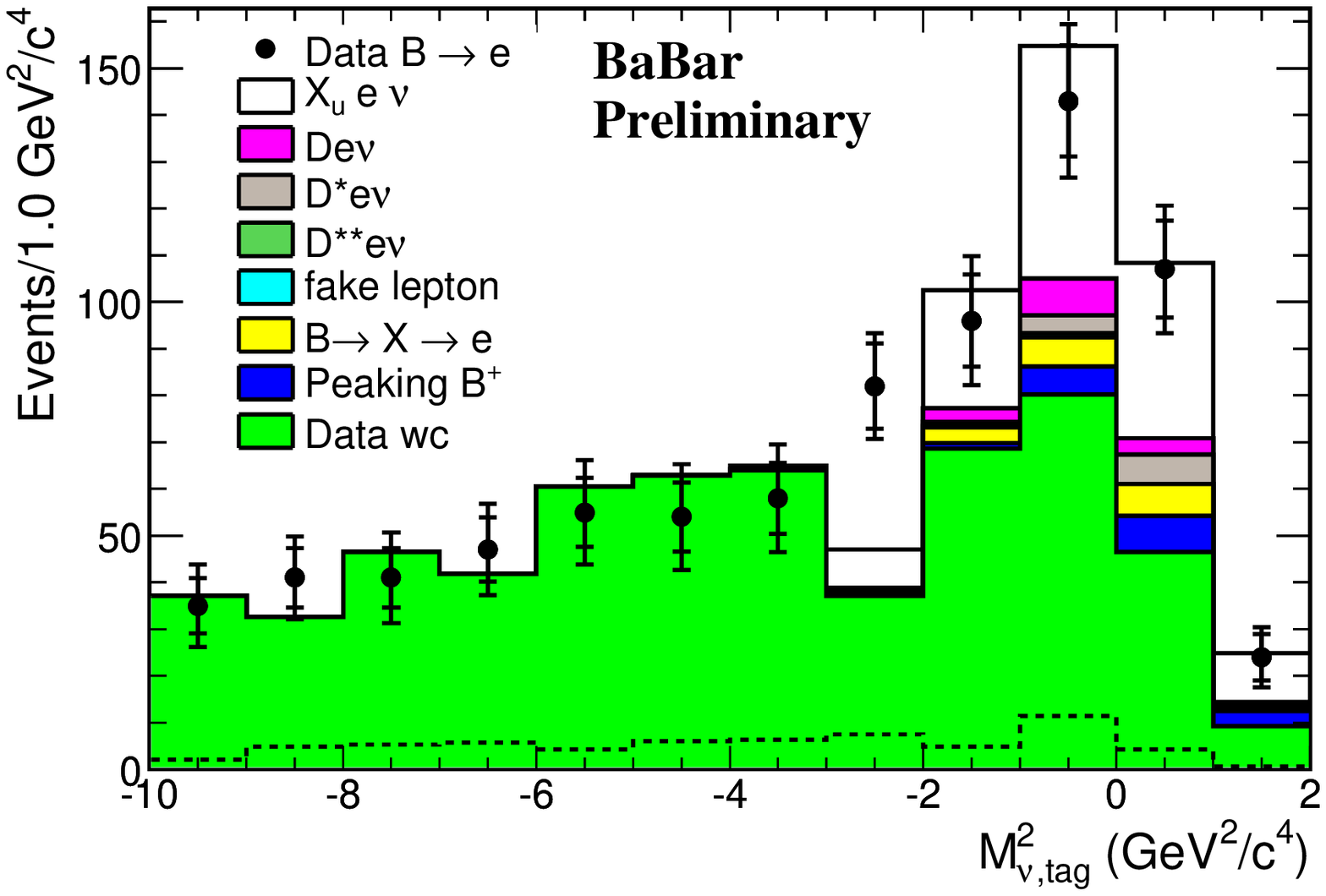}&
\includegraphics[width=8cm]{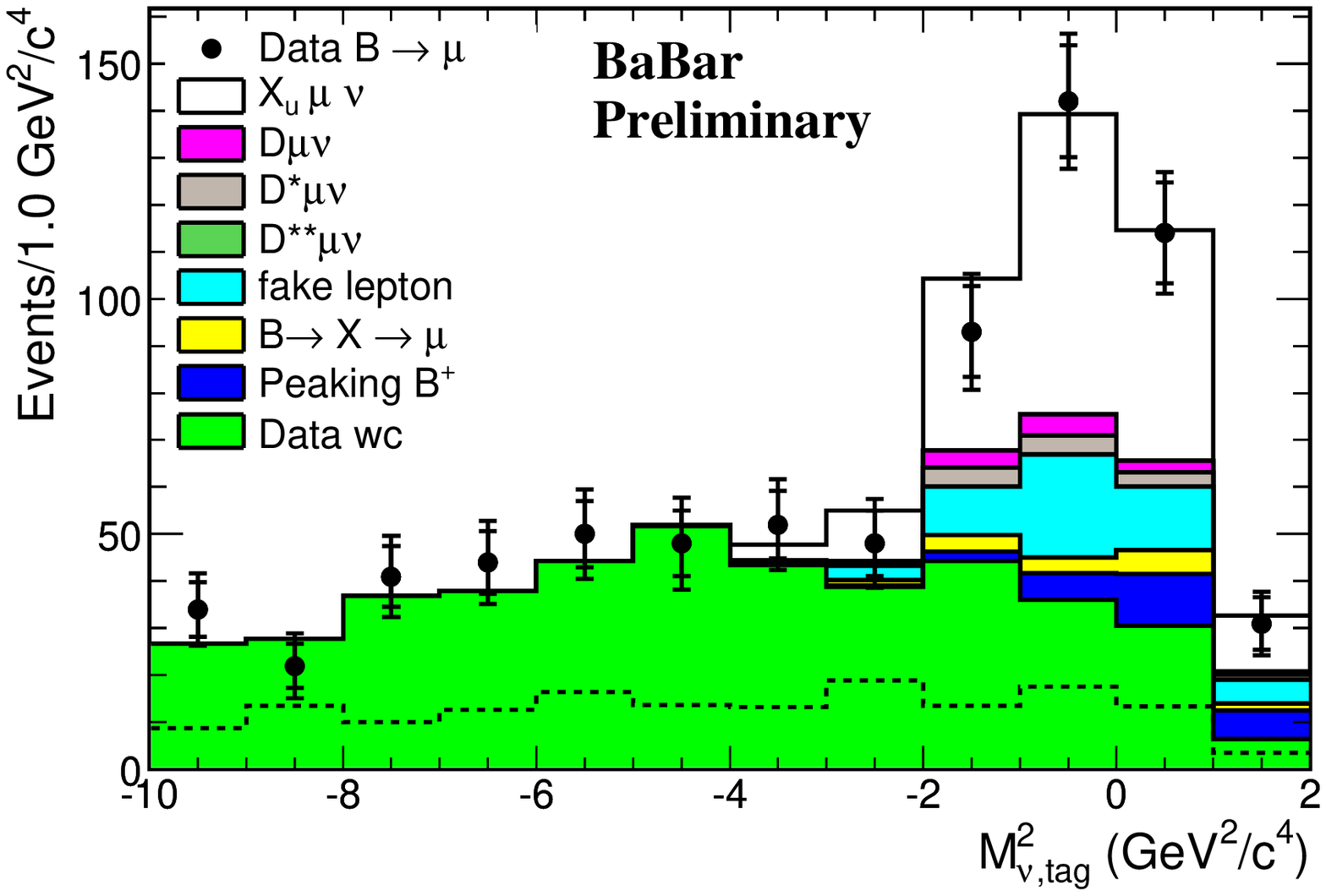}\\
\includegraphics[width=8cm]{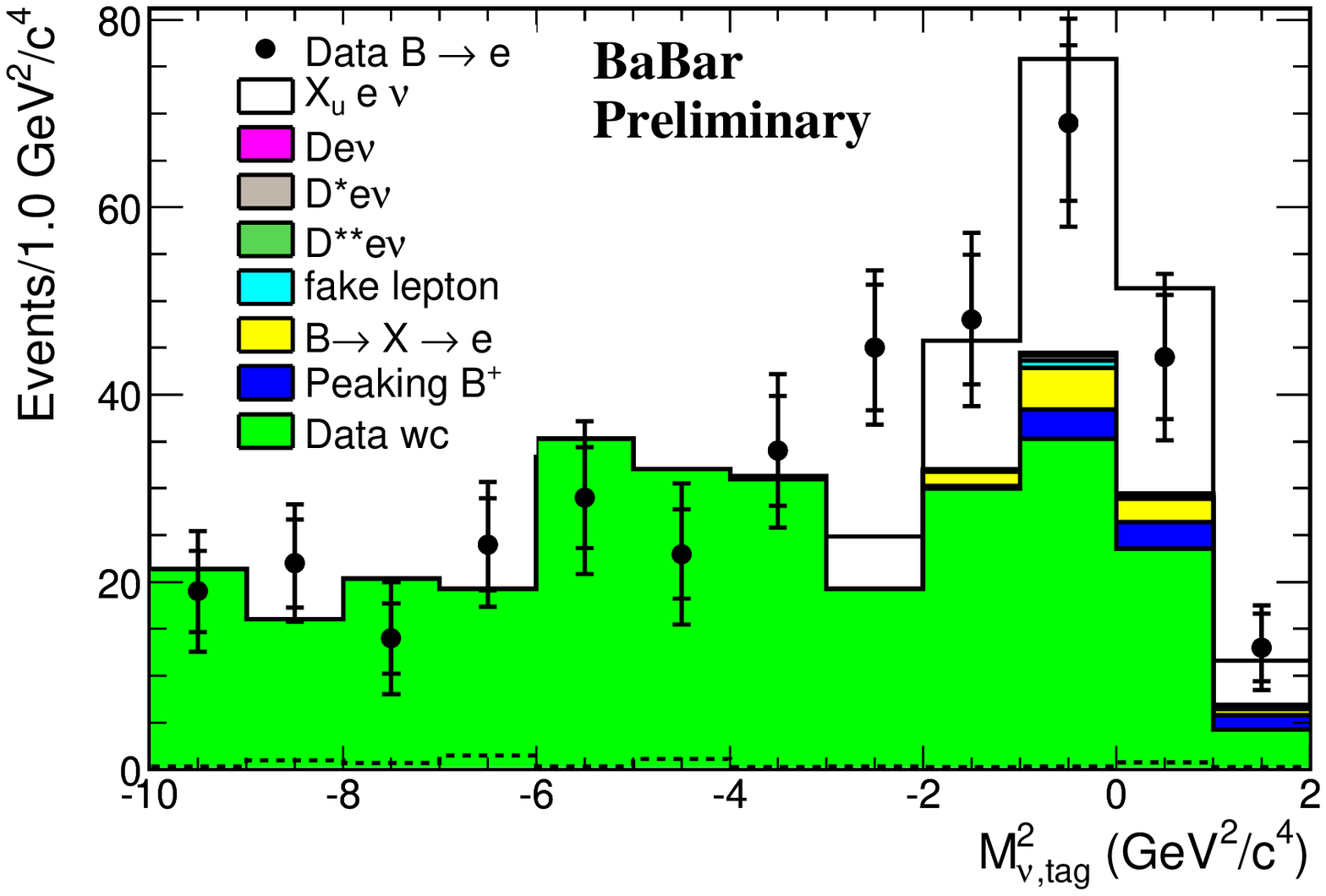}&
\includegraphics[width=8cm]{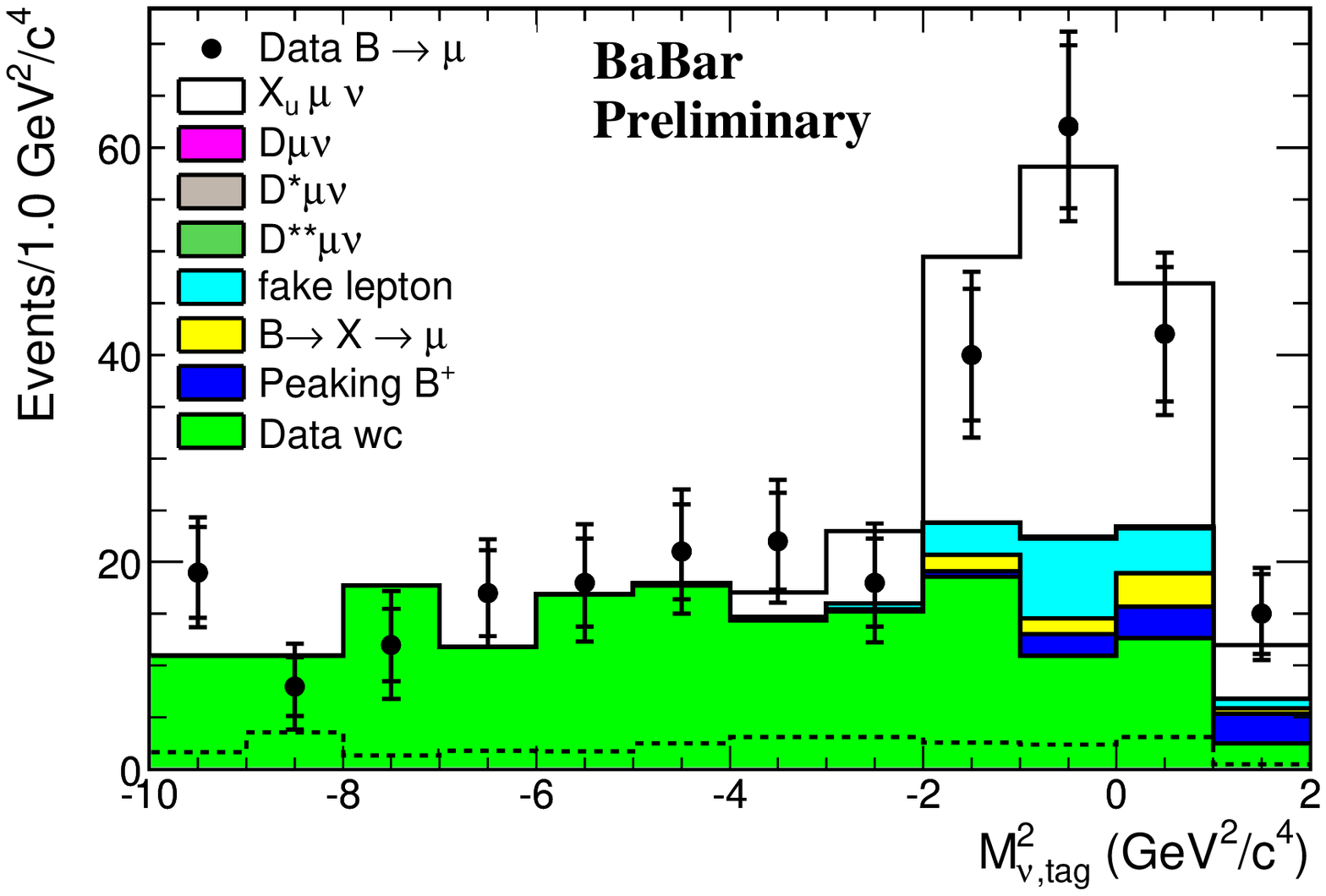}\\
\end{tabular} 
\caption{ $\mnusq_{,tag}$ distribution for $2.2<P_{\ell}<2.6\gevc$(top), $2.3<P_{\ell}<2.6\gevc$ (center) and 
$2.4<P_{\ell}<2.6\gevc$ (bottom), for $e$ (left) and $\mu$ (right). 
The signal component from simulation and the wrong-charge sample 
have been rescaled according to the fit results. The inner error bars are the
statistical error from the right-charge sample only while the larger error bars 
include also the statistical errors of the 
wrong-charge sample and of the various peaking components described by the simulation.
The distribution of the combinatorial \BB\-background, (dashed histogram) is overlaid 
to illustrate the contributions from continuum and 
non-peaking \BB\ backgrounds as expected from simulation.}
\label{f:mnufit}
\end{figure*}

   
The inclusive partial branching fraction, for a given interval $\Delta P_\ell$ in the lepton momentum, 
is calculated according to the following formula:
\begin{equation}
\Delta {\cal B}(\Delta P_\ell)=\frac{{\cal N}_u}{\varepsilon(\Delta P_\ell)\cdot {\cal N}_{\Bzb}} [1+\delta_{rad}(\Delta P_\ell)],
\end{equation}
where ${\cal N}_{u}$ is the number of fitted $B^0\to X_u\ell\nu$ events,
$\varepsilon(\Delta P_\ell)$ is the average efficiency to select 
$\Bz \to  \X_u\ell^+ \bar{\nu_\ell}$ decays in the momentum range considered,
${\cal N}_{\Bzb}$ is the total number of tagged $\Bzb$, and
$\delta_{rad}$ is a factor that corrects for the impact of final state radiation,
primarily on the lepton spectrum.
We use the simulation to compute the efficiency of the selection $\varepsilon(\Delta P_\ell)$. 
We estimate $\delta_{rad}$ by comparing the spectra generated with and without photon radiation,
simulated with {\sc PHOTOS} \cite{photos}.

The final result for the partial branching fraction for each 
interval $\Delta P_\ell$, obtained by averaging the partial branching fractions
for $e$ and $\mu$, are reported in Table~\ref{t:final} (last row).  
The average is computed with the COMBOS package 
\cite{COMBOS} and accounts for the correlations between systematic
uncertainties, which are described in the next section.

  
\section{SYSTEMATIC STUDIES}

We compute the total systematic error by adding in quadrature all uncertainties 
induced by background subtraction and efficiency calculation.

We vary the fraction of $B^+$ peaking background by $\pm 10\%$, corresponding to the uncertainty we 
assign to the isospin symmetry relation (see above).

To estimate the systematic error associated with the uncertainty in the average energies of
the colliding beams, we change in our simulation the beam energy by $1.5~\mev$ and study the 
changes in the shape of the lepton momentum spectra for the signal events.

We vary the efficiencies for $e$ and $\mu$ identification within their uncertainties by
$\pm 1.4 \%$ and $\pm 2.2\%$ respectively.

\begin{table*}
\begin{center}
\caption{\label{t:syst}Breakdown of all the sources of error (expressed in $\%$). 
The first row shows the confidence level of the e-$\mu$ averaged result, including all systematic
uncertainties.}
\begin{tabular}{l|c|c|c}
\hline
$\Delta P_\ell$ lepton momentum range  & 2.2-2.6& 2.3-2.6 & 2.4-2.6 \\
\hline
\hline 
{\bf Statistical}		&12.6	&16.1	&19.3	\\
\hline
{\bf Systematics} 	&6.1	&5.	&6.4	\\
\hline
Monte Carlo statistics  &2.8	&3.4	&5.0	\\
Peaking $B^+$           &2.5   &1.1      &1.2   \\
$N_B^0$			&1.3	&1.3	&1.3	\\
$B$ movement            &0.4   &1.0     &1.5  \\  
Event Selection		&1.0	&1.1	&1.9	\\
PID			&1.3	&1.5	&1.4	\\
Radiation		&1.1	&1.2	&1.3	\\
$J/\psi,\psi'$ bkg	&0.5	&0.2	&0.2	\\
Fake lepton		&2.3	&2.9    &1.7	\\	
$B\to D\ell\nu$		&1.9	&0.0	&0.0	\\
$B\to D^*\ell\nu$   	&2.4	&0.5	&0.0	\\
$B\to D^{**}\ell\nu$	&0.1	&0.0	&0.0   \\
$X_{u}$ composition  	&0.7	&0.4	&0.4   \\
$\ssbar$ pair production&1.2    &0.5   &0.3   \\
\hline
\end{tabular}
\end{center}
\end{table*}

The uncertainty on track reconstruction ($\pm 0.5\%$ per track) directly affects
the lepton track reconstruction. In addition,
the uncertainties in the reconstruction of the remaining charged tracks and the photons
($\pm 1.8\%$ per shower) 
introduce an error on the event selection efficiency, because they affect some of the variables
used to subtract the continuum: $N_{track}$, $p_{miss}$, $M_{tot}$ and the aplanarity$~A$. 

Misidentification rates are measured
with data control samples of $\pi$, $K$ and $p$ from $K_s,~ \Dz$, and $\Lambda$ decays.
They are less than 0.1$\%$ for electrons and about 2$\%$ for muons.
The contribution of the fake electrons to the peaking background is negligible, whereas 
there is a sizeable contamination
from fake muons (see Tab.\ref{t:final}). We vary 
the electron fake rate by $\pm 50\%$.
The uncertainty on the muon fake rate is due the to systematic error in the measurement 
of the mis-identification probabilities in data and simulation; and in the
differences in the production rate of high-momentum pions in data and in simulation.
The first effect is estimated by comparing the data-simulation correction of 
two different hadron control samples: $D^*\to \pi_{soft}(K\pi)$ and 
$\tau\to 3~prongs$. To estimate the effect of pion production, we compare 
in data and Monte Carlo the number of candidate pion tracks that fail 
both a loose muon and electron selection. 
The total uncertainty on the muon fake rate is $15\%$, that is obtained by adding in 
quadrature the two uncertainties described above. 

A sizeable source of background is due to
$B\to J/\psi(\to\ell\ell) X $ decays that are not identified because
one of the leptons is undetected. We vary their number by the error on 
the inclusive branching fraction for  $B\to J/\psi X$ decays ($\pm 3\%$)\cite{PDG}.
The contribution from $B\to \psi'(\to\ell\ell) X $ is negligible.

We model $B\to D \ell \nu $ and $B\to D^*\ell \nu$ decays with a 
parametrization of their form factors inspired by the Heavy Quark Effective Theory \cite{CLN}.
We use the latest measurements of the form-factor parameters \cite{FF} and vary each 
parameter by its error. 

We vary the values of the branching fractions for ${\cal B}(B\to D^{*} \ell \nu$) and 
${\cal B}(B \to D \ell \nu)$ 
by their errors; we assume $\pm 100\%$ systematic uncertainty on  ${\cal B}(B \to \dstrstr\ \ell
\nu)$. The amount of the remaining backgrounds has negligible effect on the fit result.
 
We study the sensitivity of the selection efficiency to the composition of the signal.
We vary the branching fractions of the various exclusive $B^0\to X_u\ell\nu$ 
decays channels within their uncertainties: $\pm 12\%$ for $B^0\to \pi\ell\nu$, 
$\pm 22\%$ for $B^0\to \rho\ell\nu$ and $\pm 14\%$ for inclusive $B^0\to X_{u}\ell\nu$.
The efficiency varies by less than one percent, being mostly sensitive to 
$B^0\to \rho\ell\nu$ decays.

The kaon veto also affects the signal efficiency due to $\ssbar$ pair production 
(also ``$\ssbar$ popping'') in semileptonic charmless $B$ decays. If we vary the kaon production
by $\pm30\%$, the result changes by $1.3\%$.
Table~\ref{t:syst} shows the relative systematic errors induced by all the sources described above.

  
\section{RESULTS AND CONCLUSIONS}

We have measured the partial branching fraction for charmless semileptonic $B^0$ 
decays for several overlapping intervals in the lepton momentum. 
The results are listed in Table \ref{t:result}. For the momentum range from 2.3 to 2.6~\gevc 
we obtain $\Delta{\cal B}(B^0\to X_u\ell\nu)=(1.30\pm0.21\pm0.07)\times 10^{-4}$. 
This first measurement for neutral $B$ mesons can be used to test isospin invariance, using the ratio

\begin{equation}
R^{+/0} = \frac{\Delta \Gamma^+}{\Delta \Gamma^0}= \frac{\tau^0}{\tau^+} \cdot \frac{\Delta{\cal B}(\Bp\to\X_u\ell\nu)} {\Delta{\cal B}(\Bz\to\X_u\ell\nu)},
\end{equation}

\noindent where ${\tau^+}/{\tau^0}=1.071\pm 0.009$ \cite{PDG} 
is the ratio of the lifetimes for $B^+$ and $B^0$. 
Since no measurement of the partial decay rate is available for charged $B$ mesons, 
we use the earlier untagged 
\babar\ measurement \cite{bad1047} for the sum of charged and neutral $B$ mesons and determine $R^{+/0}$ 
from the following expression, 

\begin{equation}
R^{+/0} = \frac{\tau^0}{\tau^+}\cdot \frac{1}{1-f_{00}} \cdot  [\frac{\Delta{\cal B}(B)}{\Delta{\cal B}(B^0)}-f_{00}],
\end{equation}

\noindent where $f_{00}=0.494\pm0.008$ \cite{PDG} is the $\Upsilon(4S)\to\B^0{\overline B}^0$ 
branching fraction.

The overlap of the data sample of the two \babar\ measurements is negligible.  
We consider the systematic errors 
due to PID efficiency and fake rates, charged particle tracking, 
$\psi$ background and radiative effect fully correlated. 
For the interval 2.3 to 2.6~\gevc, we obtain $R^{+/0} = 1.18 \pm 0.35 \pm 0.17$, compatible 1.0. We can 
also express the result in terms of the charge asymmetry, 

\begin{equation} 
A^{+/0}=\frac{\Delta \Gamma^+-\Delta \Gamma^0}{\Delta \Gamma^++\Delta \Gamma^0}=\frac{R^{+/0}-1}{R^{+/0}+1},  
\end{equation}

\noindent and obtain, for the same momentum interval, $A^{+/0} = 0.08 \pm 0.15 \pm 0.08$, 
compatible with zero. 
\begin{table*}[!b]
\caption{\label{t:result} Results of this analysis (3$^{rd}$ column), 
compared to the untagged \babar\ result (2$^{nd}$) \cite{bad1047}. The result for $R^{+/0}$ and 
$A^{+/0}$ are also reported.}
\begin{tabular}{c|r|r|r|r} 
\hline
$\Delta P_\ell$   & $\Delta{\cal B}(B)\cdot 10^4 \cite{bad1047}$    &$\Delta{\cal B}(\Bz) \cdot 10^4$  & $R^{+/0}$ & $A^{+/0}$\\
            \hline
            $2.2-2.6\gevc$ & 2.31$\pm$0.10$\pm$0.18 &  2.62$\pm$0.33$\pm$0.16 & 0.71$\pm$0.22$\pm$0.16 &-0.17$\pm$0.15$\pm$0.11\\
            $2.3-2.6\gevc$ & 1.46$\pm$0.06$\pm$0.10 &  1.30$\pm$0.21$\pm$0.07 & 1.18$\pm$0.35$\pm$0.17 & 0.08$\pm$0.15$\pm$0.08\\
            $2.4-2.6\gevc$ & 0.75$\pm$0.04$\pm$0.06 &  0.76$\pm$0.15$\pm$0.05 &  0.91$\pm$0.37$\pm$0.18 &-0.05$\pm$0.20$\pm$0.10\\
\hline
\end{tabular}
\end{table*}
Thus, with the presently available data sample,  there is no evidence for a difference 
in partial decay rates between $B^0$ and $B^+$ at the high end of the lepton momentum spectrum, 
where we would expect the impact of weak annihilation in $B^+$ decays. We set an upper limit on the 
absolute value of the charge asymmetry to $|A^{+/0}|<0.35$ at $90\%$ confidence limits (C.L.).
If we define $\Delta\Gamma_{WA}=\Delta \Gamma^+-\Delta \Gamma^0$ the contribution of the  
weak annihilation, we write the relation 

\begin{equation} 
A^{+/0}=\frac{\Delta \Gamma^+-\Delta \Gamma^0}{\Delta \Gamma^++\Delta \Gamma^0}=\frac{f_{WA}(\Delta p)\Gamma_{WA}}{2\cdot f_{u}(\Delta p)\Gamma_{u}},  
\end{equation}

\noindent where $f_{WA}(\Delta P_\ell)$ refers to the fraction of the 
weak annihilation rate contributing in the momentum interval $\Delta P_\ell$, $f_u(\Delta p_\ell)$
is the fraction of lepton spectrum in the same momentum interval $\Delta P_\ell$ and $\Gamma_{u}$ is the
total $B\to\X_u\ell\nu$ decays width. We can write the relative impact of the $\Gamma_{WA}$ on the 
$B\to X_{u} \ell {\nu_\ell}~$ as

\begin{equation} 
\frac{|\Gamma_{WA}|}{\Gamma_{u}}=\frac{2\cdot f_u(\Delta P_\ell)}{f_{WA}(\Delta P_\ell)}\cdot A^{+/0}. 
\end{equation}
 
\noindent Using $f_{u}(2.3-2.6)\approx 5.5\%$ ~\cite{blnp}, we can place, depending on $f_{WA}$, 
a limit of
\begin{equation} 
\frac{|\Gamma_{WA}|}{\Gamma_{u}} < \frac{3.8~\%}{f_{WA}(2.3-2.6)},{~~~\rm~at~90\% ~C.L.}. 
\end{equation}
This results is also consistent with a limit set by the 
CLEO Collaboration \cite{CLEO}.      


\section{ACKNOWLEDGMENTS}

\input{acknowledgements}


\end{document}

%% file: authors_lp2007.tex
\begin{center}
\small

The \babar\ Collaboration,
\bigskip

%
{B.~Aubert,}
{M.~Bona,}
{D.~Boutigny,}
{Y.~Karyotakis,}
{J.~P.~Lees,}
{V.~Poireau,}
{X.~Prudent,}
{V.~Tisserand,}
{A.~Zghiche}
\inst{Laboratoire de Physique des Particules, IN2P3/CNRS et Universit\'e de Savoie, F-74941 Annecy-Le-Vieux, France }
{J.~Garra~Tico,}
{E.~Grauges}
\inst{Universitat de Barcelona, Facultat de Fisica, Departament ECM, E-08028 Barcelona, Spain }
% ...

\end{center}\newpage

%% file: acknowledgements.tex
We are grateful for the 
extraordinary contributions of our \pep2\ colleagues in
achieving the excellent luminosity and machine conditions
that have made this work possible.
The success of this project also relies critically on the 
expertise and dedication of the computing organizations that 
support \babar.
The collaborating institutions wish to thank 
SLAC for its support and the kind hospitality extended to them. 
This work is supported by the
US Department of Energy
and National Science Foundation, the
Natural Sciences and Engineering Research Council (Canada),
Institute of High Energy Physics (China), the
Commissariat \`a l'Energie Atomique and
Institut National de Physique Nucl\'eaire et de Physique des Particules
(France), the
Bundesministerium f\"ur Bildung und Forschung and
Deutsche Forschungsgemeinschaft
(Germany), the
Istituto Nazionale di Fisica Nucleare (Italy),
the Foundation for Fundamental Research on Matter (The Netherlands),
the Research Council of Norway, the
Ministry of Science and Technology of the Russian Federation, and the
Particle Physics and Astronomy Research Council (United Kingdom). 
Individuals have received support from 
CONACyT (Mexico),
the A. P. Sloan Foundation, 
the Research Corporation,
and the Alexander von Humboldt Foundation.